# Summer Mesosphere Temperature Distribution from Wide-Angle Polarization Measurements of the Twilight Sky

Oleg S. Ugolnikov*[1], Igor A. Maslov[1,2]

[1]Space Research Institute, Russian Academy of Sciences,
Profsoyuznaya st., 84/32, Moscow, 117997, Russia
[2]Moscow State University, Sternberg Astronomical Institute,
Universitetskii prosp., 13, Moscow, 119991, Russia
ougolnikov@gmail.com, imaslov@iki.rssi.ru.
*Corresponding author, phone: +7-916-391-73-00, fax: +7-495-333-51-78.

**Abstract**

The paper contains the results of wide-angle polarization camera (WAPC) measurements of the twilight sky background conducted in summer 2011 and 2012 at 55.2°N, 37.5°E, southwards from Moscow. The method of single scattering separation based on polarization data is suggested. The obtained components of scattering matrixes show the domination of Rayleigh scattering in the mesosphere for all observation days. It made possible to retrieve the altitude distribution of temperature in the mesosphere. The results are compared with the temperature data by TIMED/SABER and EOS Aura/MLS instruments for nearby dates and locations.

**1. Introduction.**

Thermal properties of upper atmosphere were the object of interest for a long time, but there was no way to investigate them until the rocketsonde epoch in the middle of XX century. First attempts to measure the temperature in the layers referred to the mesosphere now were made earlier basing on the meteors moderation observations. Lindemann and Dobson (1923) had found that the upper atmosphere (60 km) is heater than the lower stratosphere, where temperature was already known. After the method modification by Fedynski and Stanukovich (1935) and Whipple (1943) it was shown that the temperature decreases at 80 km, but its values were sufficiently overestimated.

However, the appearance of polar mesospheric clouds in late XIX century and their regular observations after that revealed the relation with mesosphere thermal regime (Balsley and Huaman, 1997). This shows the global process of changing of physical conditions in mesosphere of the Earth during the last decades. The reason of such changes can be related with increasing contribution of trace gases in all atmosphere layers. Radiative cooling by $CO_2$ was found to be one of the basic mesosphere cooling mechanisms causing the summer temperature minimum in mid-latitudes (Kuhn and London, 1969).

Investigation of such relation required the systematic mesosphere temperature measurements. The first long-time temperature data obtained by Kokin and Lysenko (1994) had shown sufficient trend value: about –1K per year. Further observations and analysis reviewed by Beig et al. (2003) reduced this amount and showed its latitude dependency.

The basic problem of mesospheric investigations is difficulty of measurements, especially *in situ* (Dyrland et al., 2010). The mesosphere is too high for balloons but too low for the spacecrafts. During high-latitude summer period when the extremely low temperatures are observed, the mesosphere is continuously illuminated by the Sun, that restricts the time and location area of some types of measurements, such are nightglow emission and twilight observations. The present experimental methods are described in details by Lübken et al. (1994) and Beig et al. (2003). They can be separated in three groups: rocketsonde *in situ* measurements, ground-based and spacecraft remote sensing. Optical ground-based sensing is basically held by lidars. Rayleigh lidar techniques (Hauchecorne and Chanin, 1980) scan the backscattering coefficient depending on the altitude which can be the basis of the temperature calculation if the contribution of aerosol scattering is small. Sodium (She at al., 1992) and potassium (von Zahn and Hoffner, 1996) lidars retrieve the temperature profiles inside the Na and K layers with better accuracy owing to higher S/N ratio.

First continuous spaceborn mesosphere temperature data were obtained by NASA Solar Mesosphere Explorer satellite (SME) in 1980s by measurements of Rayleigh scattering (Clancy and Rusch, 1989). The same method was used by Wind Imaging Interferometer (WINDII) onboard the NASA Upper Atmosphere Research Satellite (UARS, Reber et al., 1993) operated in 1991-2005. The wavelength of WINDII analysis (553 nm, Shepherd et al., 2001) is close to the one used in present paper. Mesosphere temperature values were also obtained by Halogen Occultation Experiment (HALOE, Remsberg et al., 2002) and High Resolution Doppler Imager (HRDI, Thulasiraman and Nee, 2002) onboard the same satellite.

Recently, the mesosphere temperature is measured by the Sounding of the Atmosphere Using Broadband Emission Radiometry instrument (SABER, Russell et al., 1999) onboard the Thermosphere, Ionosphere, Mesosphere, Energetics and Dynamics (TIMED) satellite launched in 2001, and Microwave Limb Sounder (MLS, Schwartz et al., 2008) onboard Earth Observing System (EOS) Aura satellite, launched in 2004, both by NASA. The experiments are based on the measurements of microwave emission of carbon dioxide (SABER) and oxygen (MLS). SABER data shows rapid changes of summer mesosphere temperature at the latitudes above 40-50° (Xu et al., 2007) which makes all kinds of measurements especially important there. Polar mesopause summer temperatures by MLS and meteor radar data (Dyrland et al., 2010) are just 10-20K less.

Twilight method of atmosphere sounding was under consideration for a long time after the pioneer work by Fesenkov (1923). But during the following decades the possibilities of the method were restricted by the contribution of multiple scattering, that was not known exactly even for the light twilight and sufficiently increased during the darker stage. Wide review of observational and theoretical works by Rozenberg (1966) shows the underestimation of this contribution. Fast sky depolarization always observed at solar zenith angles from 95° to 99° (transitive twilight period by classification of Ugolnikov and Maslov (2007)) and caused by multiple scattering but was often related with upper atmospheric aerosol that led to its large concentration and unreal optical properties. That gave no possibility to separate the single Rayleigh scattering and to measure the temperature.

Experimental method of single scattering separation with account of polarization was suggested by Fesenkov (1966) and improved by Ugolnikov (1999), Ugolnikov and Maslov (2002). The results were in good agreement with the numerical simulations (Ugolnikov et al., 2004) during the light twilight, but dark stage remains hard for exact computer modeling (Postylyakov, 2004) even in a present time due to fast increase of scattering order. The basic experimental problem of mesosphere research by transitive and dark twilight analysis was the necessity of fast simultaneous measurements of different sky points during the short period. Using the scanning photometers (Zaginailo, 1993, Ugolnikov, 1999) or first generation CCDs (Ugolnikov and Maslov, 2002) provided just the small numbers of measuring acts corresponding to the mesosphere layer. However, the accuracy of polarization measurements had increased, and it helped to detect the depolarization effects caused by meteoric dust in the mesosphere (Ugolnikov and Maslov 2007). In this paper we present the results of the measurements that held frequently and simultaneously over the sufficient part of celestial hemisphere.

**2. Observations.**

WAPC measurements had started in early summer of 2011 at the point 55.2°N, 37.5°E, about 60 km southwards from Moscow. From one side, this latitude is high enough to observe the summer mesosphere cooling. From another side, the twilight (even in June) is quite deep to hold the single and multiple scattering separation procedure described below and explore the mesosphere single scattering. The observation location is near the northern border of the area where it can be done. The mesopause moves down to 85 km during the summer here (Xu et al., 2007), this altitude is in the work range of twilight sounding method.

The results of very first observation dates are presented by Ugolnikov and Maslov (2012), initial device properties are also described there. After some primary sessions in 2011 observations continued in the late spring and summer period of 2012 with improved measuring device. It consists of three consecutive lenses providing wide field of view (almost 140°), but the light crosses the optical axis by a small angle (not more than 10° for the edge of the frame) in the place where the polarization filter is installed. This filter makes 120°-steps once in 40 seconds during the light twilight and once in a minute during the dark twilight and the night. Spectral filters fix the observational band, the same for 2011 and 2012, with effective wavelength equal to 540 nm. The sky images are created by CCD-matrix Meade DSI PRO II, angular resolution (1 pixel) is about 0.3°. For better accuracy the sky background data are averaged inside the circles with radius 10 px (about 3°) placed in 5° one from another.

The observations start before the sunset or during the light twilight and continue throughout the night. The exposure times vary from 0.001 to 15 seconds. Stars astrometry and photometry data in the night sky frames help to fix the camera orientation, field curvature and self polarization. The star data is also needed to define the multiplication of camera flat field and the atmospheric transparency in different sky points − this value is necessary to calculate the single scattering function. The procedure works well for zenith distances up to 60°, it is the restriction for the observational data to be processed. Analysis of transitive and dark twilight background described below is made for the moonless twilights and the twilights with the crescent Moon with the phase less than 0.5.

**3. Twilight sky polarization properties and single scattering separation.**

The initial analysis of first WAPC observations (Ugolnikov and Maslov, 2012) had included only the points of solar vertical as it was done in many papers before (Zaginailo, 1993, Ugolnikov, 1999, Ugolnikov and Maslov, 2002). Here we expand it to the whole observable part of the sky. To do it, we have to make some definitions. The point where the intensity and polarization are measured is characterized by two coordinates $\zeta$ and $\tau$ (see Fig.1). For solar vertical points the value of $\zeta$ is equal to the zenith angle of the point, with negative sign in the sky area opposite to the Sun. The value of $\tau$ is equal to zero there. The zenith angle of observation point is denoted by $z(\zeta, \tau)$, $z_0$ is the ephemerid zenith angle of the Sun (with no account of refraction). The single scattering angle $\theta(\zeta, \tau)$ is equal to the angular distance between the Sun and observation point with the small altitude-dependent correction by the atmospheric refraction.

Far from two horizon points at 90° from the Sun we may consider the Earth as cylinder with axis perpendicular to the solar vertical plane. This simplification can be used since the distance to the scattering point (not more than 200 km for $|z|<60°$) is many times less than the Earth's radius. In this model the Earth's shadow altitude is the function of $z_0$ and $\zeta$ not depending on $\tau$. Let **A** be the sky point with coordinates ($\zeta$, $\tau$), see Fig.1. The measured values are three components of Stokes vector of twilight background:

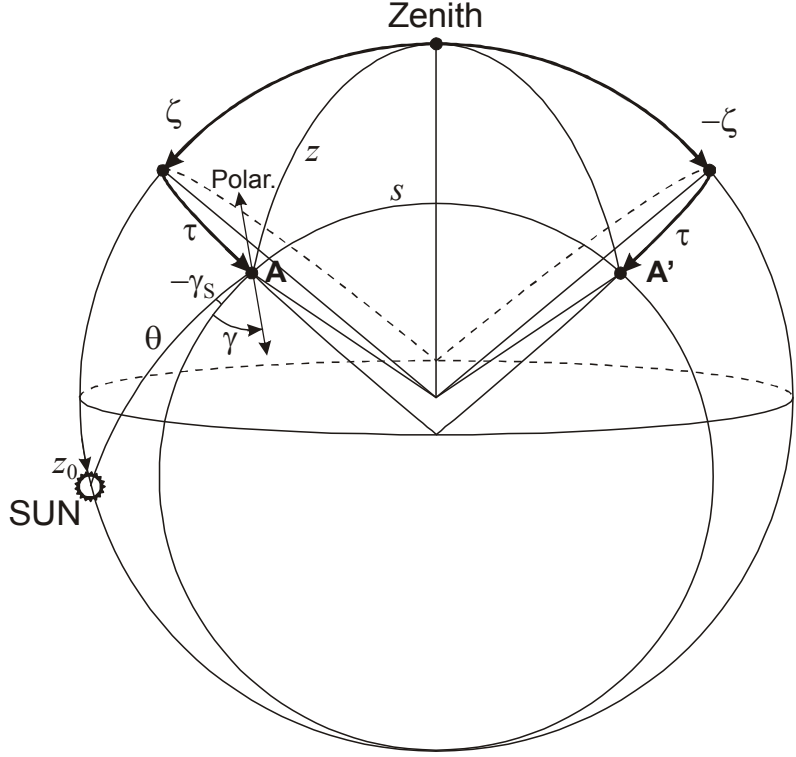

*Figure 1. Sky coordinate system used in this work.*

$$\mathbf{B}(\zeta,\tau) = \begin{pmatrix} I(\zeta,\tau) \\ I(\zeta,\tau)Q(\zeta,\tau)\cos 2\gamma(\zeta,\tau) \\ I(\zeta,\tau)Q(\zeta,\tau)\sin 2\gamma(\zeta,\tau) \end{pmatrix} \quad (1).$$

Here $I$ is the intensity, $Q$ is the polarization (such denotation is chosen in order not to mix it up with pressure value that will appear below), $\gamma$ is the angle between the polarization direction and the reference line on the celestial sphere. To have a common system providing the continuous vector function $\mathbf{B}(\zeta, \tau)$ for the whole sky above the horizon without singularities, we choose the reference line as a minor circle of the celestial sphere containing the points $\mathbf{A}(\zeta, \tau)$, $\mathbf{A}'(-\zeta, \tau)$ and nadir point. This circle is denoted by $s$ in Fig.1.

In the solar vertical ($\tau=0$) the third Stokes vector component vanishes, and the second one is equal to $\pm IQ$, the sign depends on $\zeta$ value. Fig.2 shows the dependency of the sky polarization in different solar vertical points for the morning twilight of May, 26, 2012. The analysis of polarization value behavior and its relation with the single and multiple scattering intensities is presented in (Ugolnikov and Maslov, 2007, 2012). During the dark period of twilight ($z_0$ above 100.5° for observable sky area) the polarization becomes practically symmetric relatively the zenith point. As it was shown in those papers, the multiple scattering (together with night sky component) completely dominates in the sky background during this twilight period. Single scattering appears at lower solar zenith angles, first of all, in the dawn area. This effect is seen as the increase of sky polarization breaking its symmetry. Corresponding moment is shown by the arrows in Fig.2. Analysis of same dependency for all observed twilights leads to the empirical relation between the value of $\zeta$ and the solar zenith angle of this twilight stage:

$$z_{0S} = 99.0 + 0.02 \cdot \zeta \quad (2).$$

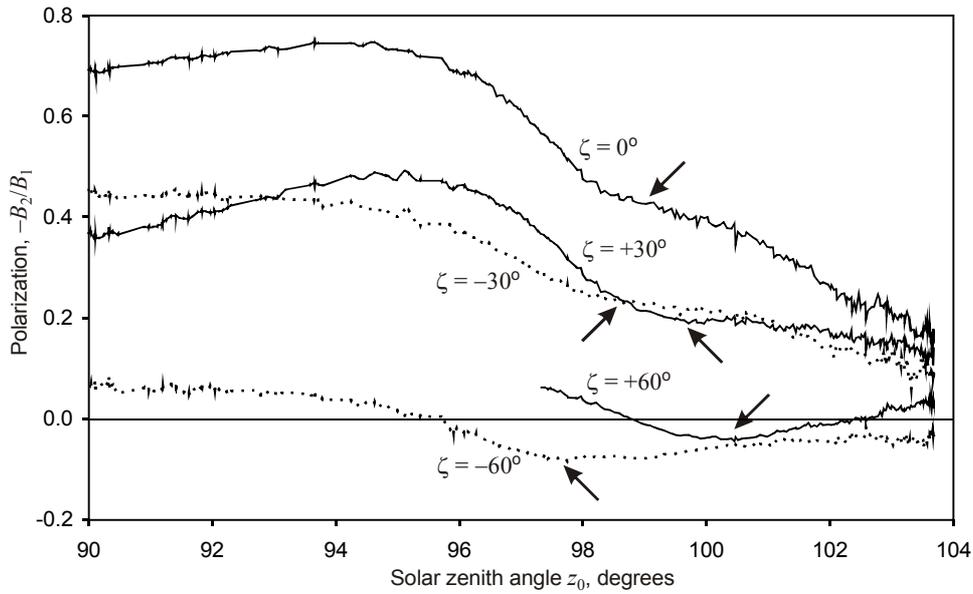

*Figure 2. The dependencies of sky polarization in different solar vertical points on a solar zenith angle (morning twilight of May, 26, 2012). Arrows show the start of total multiple scattering domination period (dark twilight). The light twilight data for +60° is absent due to partial horizon obscuration at the observation location.*

Here both values are expressed in degrees. This formula is approximate and needed only to restrict the range of multiple scattering domination. The $z_{0S}$ value is considered as the upper limit (Ugolnikov and Maslov (2007) used the zenith alue 98.5° for different season and location). The corresponding geometrical shadow altitude is about 75 km. However, strong absorption of tangent solar emission in the troposphere sufficiently increases the scattering and twilight layer baseline altitudes (see below).

The basis of single scattering separation method used by Zaginailo (1993), Ugolnikov and Maslov (2012) is the constant brightness ratio of multiple scattering in symmetrical solar vertical points ($\zeta$, 0) and ($-\zeta$, 0). Expanding the method on the polarization vector data, the assumption of constant multiple scattering polarization in definite sky point was made. Dark twilight polarization changes were related only with changing contribution of night sky background. This assumption seemed to be adequate for dawn area points with $\tau=0$ where polarization is almost constant at $z_0$ from 99.5° to 101° (see Fig.2). This period is characterized by multiple scattering domination and small night sky contribution ("dark twilight" by Ugolnikov and Maslov (2007)). However, for other sky points the polarization changes are seen. Fig.3 contains the dependencies of second and third normalized Stokes vector components ($B_2/B_1$ and $B_3/B_1$) for the points (30°, 30°) and (−30°, 30°) during the same twilight. We see the polarization changes together with small difference of polarization values in these two sky points during the dark twilight.

To improve the method of single scattering separation, we build the diagram ($B_i(\zeta, \tau) - B_i(-\zeta, \tau)$) for the points **A** and **A'** in Fig.1 during the period of multiple scattering domination ($z_0 > z_{0S}(\zeta)$). Fig.4 presents the example of this diagram for the same twilight and sky points as Fig.2. For the convenience, the scale for the first Stokes component ($B_1$) is decreased by the factor of 4 in this figure. We see that the linear correlations are absolutely clear, especially for the first vector component, but its coefficient differs for the brightness and two other Stokes vector components. It is the reflection of small difference of the sky polarization in these sky points. Basing on these correlations, we write the equation relating the multiple scattering Stokes vectors in these points:

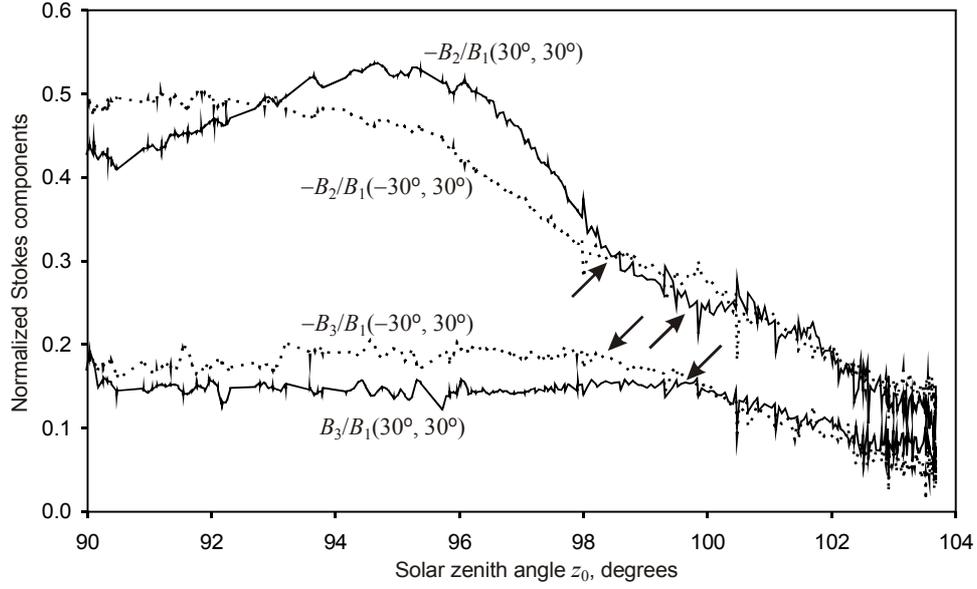

*Figure 3. The dependencies of normalized second and third Stokes components in sky points (30°, 30°) and (–30°, 30°). Twilight date and arrow denotations are the same as in Fig.2.*

$$\mathbf{B}(\zeta,\tau,z_0) = \mathbf{b}(\zeta,\tau) + \begin{pmatrix} C_I(\zeta,\tau) & 0 & 0 \\ 0 & C_Q(\zeta,\tau) & 0 \\ 0 & 0 & -C_Q(\zeta,\tau) \end{pmatrix} \times \mathbf{B}(-\zeta,\tau,z_0)$$

(3).

The vector $\mathbf{b}(\zeta,\tau)$ is contributed by the night sky background in both sky points. For transitive twilight, when $z_0 < z_{0S}(\zeta)$, the single scattering component appears in the dawn area ($\zeta > 0$). Fig.5 contains the expanded $B_1(\zeta, \tau) - B_1(-\zeta, \tau)$) diagram including this period of twilight, when the points deviate upwards from obtained line. The same effect appears for two other Stokes components. For the range $z_{0S}(-\zeta) < z_0 < z_{0S}(\zeta)$ we have the expression of the Stokes vector of single scattering:

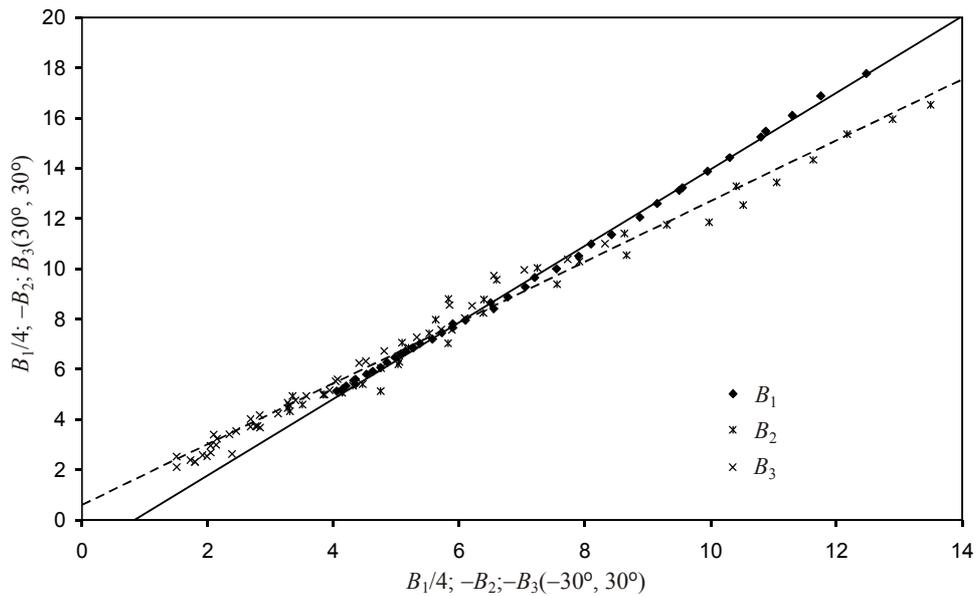

*Figure 4. Diagram of three Stokes components in sky points (30°, 30°) and (–30°, 30°) during the dark twilight. The twilight date is the same as in Fig.2.*

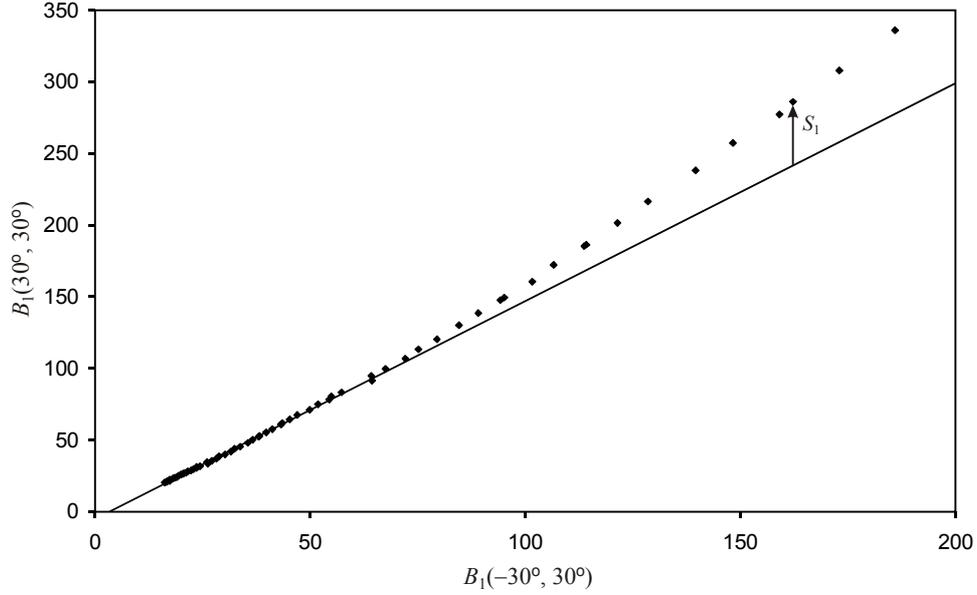

*Figure 5. Diagram of twilight background intensity in sky points (30°, 30°) and (−30°, 30°) during the transitive twilight. The twilight date is the same as in Fig.2.*

$$\mathbf{S}(\zeta,\tau,z_0) = \mathbf{B}(\zeta,\tau,z_0) - \mathbf{b}(\zeta,\tau) - \begin{pmatrix} C_I(\zeta,\tau) & 0 & 0 \\ 0 & C_Q(\zeta,\tau) & 0 \\ 0 & 0 & -C_Q(\zeta,\tau) \end{pmatrix} \times \mathbf{B}(-\zeta,\tau,z_0)$$

(4).

Here we should note that the single scattering intensity ($S_I$), most important for temperature retrieval, is calculated based only on the intensity data and well-defined constant $C_I$. Polarization data is necessary to detect the possible presence of aerosol scattering and to control the atmospheric conditions stability during the twilight. The range of analysis $z_{0S}(-\zeta) < z_0 < z_{0S}(\zeta)$ is not so wide, but it is enough to cover the middle and upper mesosphere.

**4. Single scattering functions and mesosphere temperatures.**

The principal scheme of single scattering during the deep twilight stage, when the scattering takes place above dense absorbing atmospheric layers, is shown in Fig.6. Sun is under the horizon and its emission does not reach the area below the shadow altitude $h_U$. Scattering takes place in well-illuminated atmosphere layers above the definite altitude $h_B$, and the measured value of brightness is proportional to the pressure $p$ at the altitude $h_B$. If the contribution of aerosol scattering is small relatively Rayleigh scattering, the Stokes vector of single scattering far from horizon will take the form:

$$\mathbf{S}(\zeta,\tau,z_0) = const \frac{p(h_B(\zeta,z_0))\mathbf{F}(\theta(\zeta,\tau,z_0))}{\cos z(\zeta,\tau)} \cdot E(z(\zeta,\tau)) \cdot M(z(\zeta,\tau))$$

(5).

Here $\mathbf{F}(\theta)$ is the vector scattering function per molecule, $E(z)$ is the local lower atmospheric transparency, and $M(z)$ is the camera flat field function. The multiplication $E(z) \cdot M(z)$ is known from the stars photometry data. If there were no atmospheric extinction and refraction ray divergence along the trajectory from the Sun to the scattering point, the value of $h_B$ would be equal to the shadow altitude $h_U$. In fact, this value (we call it "effective layer baseline altitude") increases. It is the character altitude immersing into shadow at the definite Sun zenith angle $z_0$ (see Fig.6). Denoting the altitude scattering profile as $J(h, \zeta, z_0)$, we define the $h_B$ altitude as:

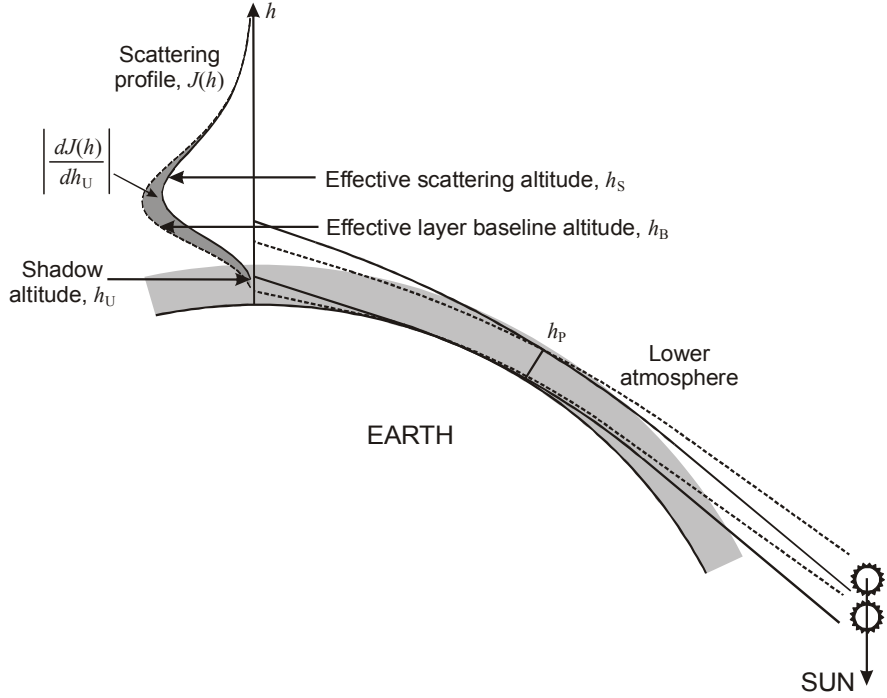

*Figure 6. Twilight optical scheme and effective altitude values definitions.*

$$h_B(\zeta, z_0) = \frac{\int_0^\infty h \cdot \left|\frac{dJ}{dh_U}\right|(h,\zeta,z_0)dh}{\int_0^\infty \left|\frac{dJ}{dh_U}\right|(h,\zeta,z_0)dh}$$

(6).

Numerical analysis for 540 nm with account of refraction effects shows that during the transitive and dark period of twilight ($z_0$ about 97°-100°) this altitude corresponds to the solar ray path with perigee altitude $h_P$ equal to 14 km (see Fig.6). This $h_P$ value is almost independent on $\zeta$ and $z_0$ during these twilight stages. Effective layer baseline altitude $h_B$ is less than the effective scattering altitude $h_S$ used by Ugolnikov and Maslov (2012), the difference is about the altitude distribution scale in the mesosphere (close to 5 km), $h_B$ value is more convenient for the pressure and temperature analysis. Fig.7 shows the dependencies $h_B(z_0)$ for different $\zeta$ values ($h_B$ and $h_U$ are independent on $\tau$ as it was discussed above). The work field of the single scattering separation method is painted in gray. Wide coverage of scattering angles and suitable accuracy are achieved for the altitudes from 70 to 85 km.

Taking the datasets for different ($\zeta$, $\tau$) points and interpolating them to the definite value of $h_B$, we obtain the vectors $p \cdot \mathbf{F}(\zeta, \tau)(h_B)$. Projecting these vectors on the line "Sun-sky point" in the sky (see Fig.1), we obtain the single scattering polarization values:

$$q(\theta(\zeta,\tau), h_B) = -\frac{(pF)_2 \cos 2\gamma_S + (pF)_3 \sin 2\gamma_S}{(pF)_1}$$

(7).

The quantity of $q$ is positive if the polarization is directed perpendicular to the scattering plane (this always is for Rayleigh scattering). Fig.8 shows the dependencies $q(\theta)$ for different altitudes for the morning twilight of May, 26, 2012 compared with the one for Rayleigh scattering in two-atomic-molecule medium:

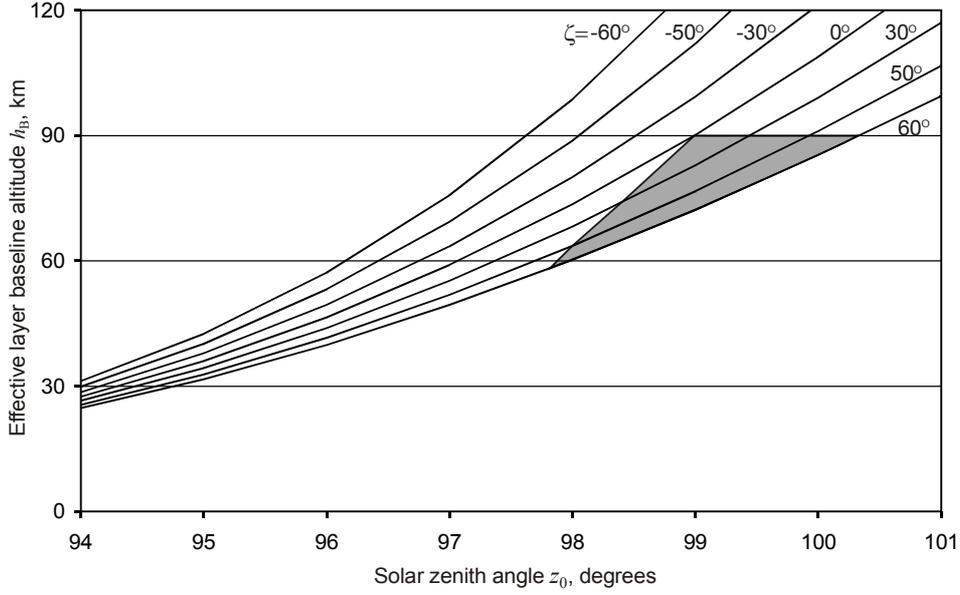

*Figure 7. Effective layer baseline altitude as a function of sky coordinate ζ and solar zenith angle. Gray triangle shows the area where single scattering appears only in dawn area of the sky.*

$$q_R(\theta) = \frac{sin^2\theta}{1.06 + cos^2\theta} \quad (8).$$

We see that the experimental polarization values are close to Rayleigh ones, depolarization possible caused by aerosol scattering is small. The same picture is also observed for other twilights. To show it, we define the general polarization characteristics $q_0$ for the whole observable range of scattering angles:

$$q(\theta, h_B) = q_0(h_B)\, q_R(\theta) \quad (9).$$

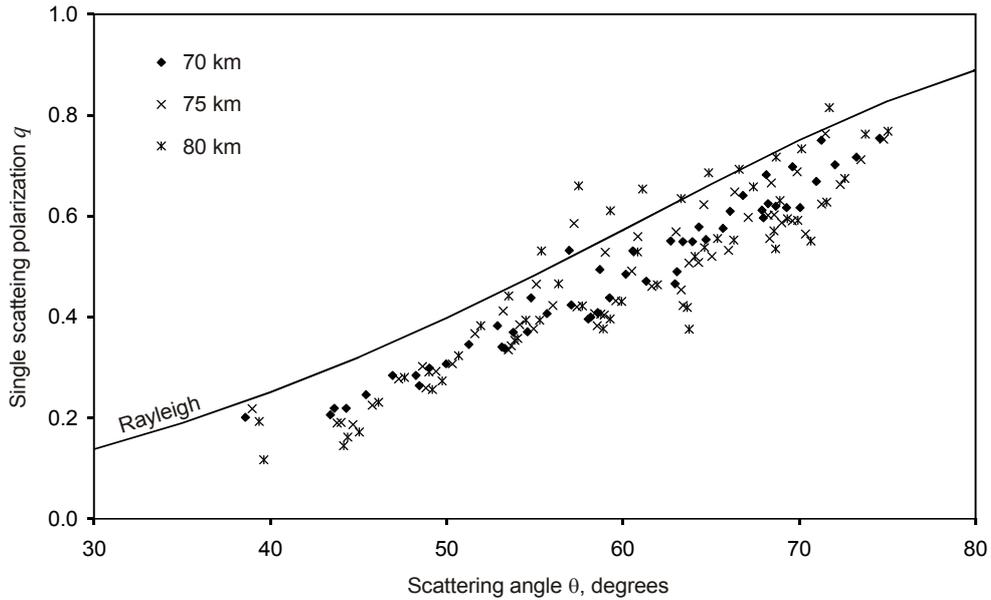

*Figure 8. Single scattering polarization depending on the scattering angle for different altitudes. The twilight date is the same as in Fig.2.*

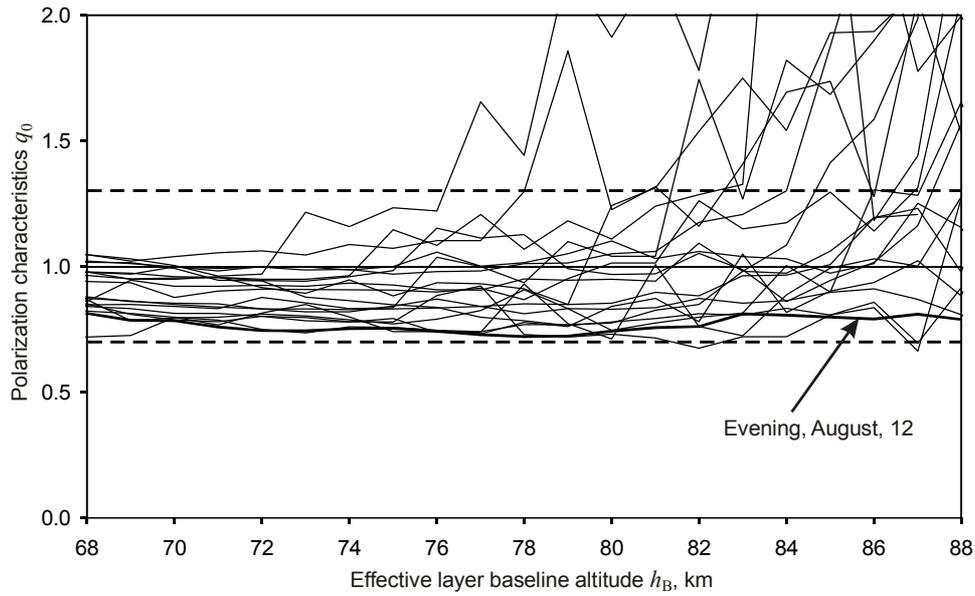

*Figure 9. Polarization characteristics of single scattering depending on the altitude for observed twilights in 2012.*

The value of $q_0$ is found by least squares method. Fig.9 shows the dependencies $q_0(h_B)$ for observational twilights in 2012. We see that the Rayleigh-like character of scattering ($q_0 \sim 1$) is a common property of mesosphere in the observation period (the 2011 picture is principally the same). Here we should add that the polar mesospheric clouds activity was not so high, they appeared just once during the observations period, below the measured sky part. Fig.9 also contains the data of evening twilight of August, 12, 2012, right after the maximum of Perseids meteor shower (solid line). It does not show sufficient depolarization, however, the $q_0$ parameter during this twilight was minimal over the whole 2012 observation set.

Larger positive deviations are seen for some twilights above 80 km, they related with less accuracy of polarization measurements and possible lower atmosphere conditions changes, having an influence to the single scattering separation procedure. The points with $q_0<0.7$ and $q_0>1.3$ (dashed lines in Fig.9) are excluded from further analysis. Thus, the polarization measurements help to separate data with stable weather conditions and Rayleigh-dominated scattering for the temperature analysis. For the data passed this criterion, the $(pF)_1$ values can be approximated by Rayleigh scattering functions:

$$(pF)_1(\theta(\zeta,\tau),h_B) = p(h_B) \cdot F_{1R}(\theta) = p(h_B)(1.06 + cos^2\theta) \qquad (10).$$

The examples of such approximation for the morning twilight of May, 26, 2012 are shown in Fig.10. We see very good agreement of measured scattering functions with Rayleigh one, confirming the small contribution of aerosol. This analysis allows to calculate the pressure values $p(h_B)$ in arbitrary units by least squares method, that is doing on 1-km grid. The logarithmic derivative of the pressure value is equal to:

$$\frac{d \ln p}{dh_B} = \frac{dp}{p dh_B} = -\frac{\mu g(h_B)}{R_G T(h_B)} \qquad (11).$$

Here $\mu$ is the molar mass, $g$ is the gravitational acceleration, $R_G$ is the gas constant. This way the temperature $T(h_B)$ can be estimated.

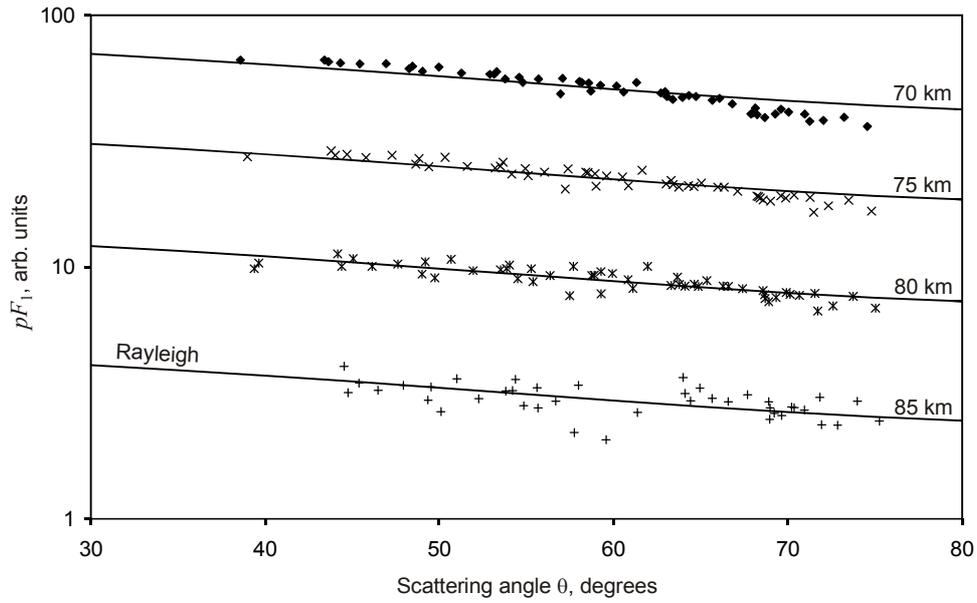

*Figure 10. Single scattering functions at different altitudes. Twilight date is the same as in Fig.2.*

Fig.11 shows the temperature values for 70, 75, 80, and 85 km for observed twilights in 2011 and 2012 with errors less than 10K. These results are compared with 2012 TIMED/SABER and EOS Aura/MLS measurements. Spaceborn data interpolated to the same altitude are averaged by 8-days interval, including the locations with latitude ±3° and longitude ±10° away from the observations place; 1σ-interval is shown. SABER data for northern mid-latitudes is available only for late spring and early summer of both years, the instrument was switched to southern hemisphere after that. The difference between 2011 and 2012 is not more than 5K.

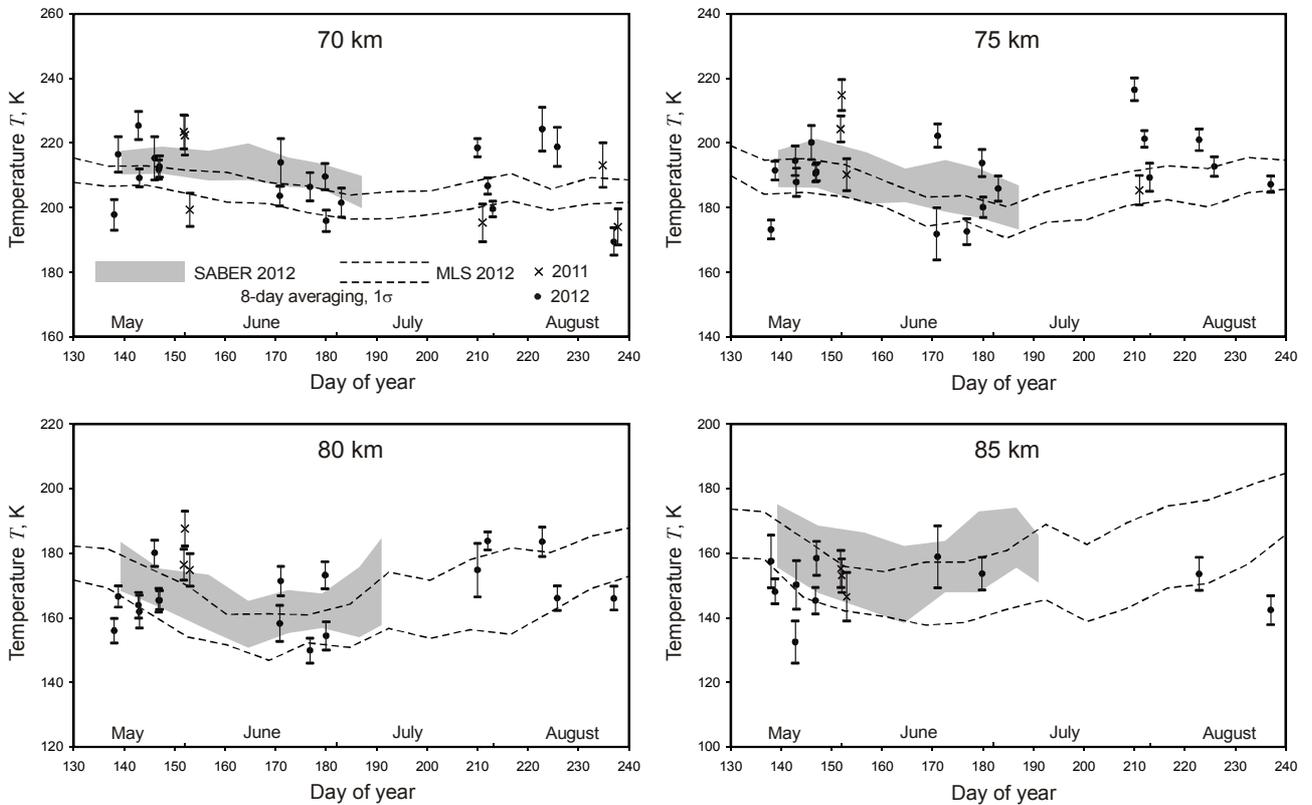

*Figure 11. WAPC temperature values in 2011 and 2012 compared with averaged SABER and MLS data in 2012.*

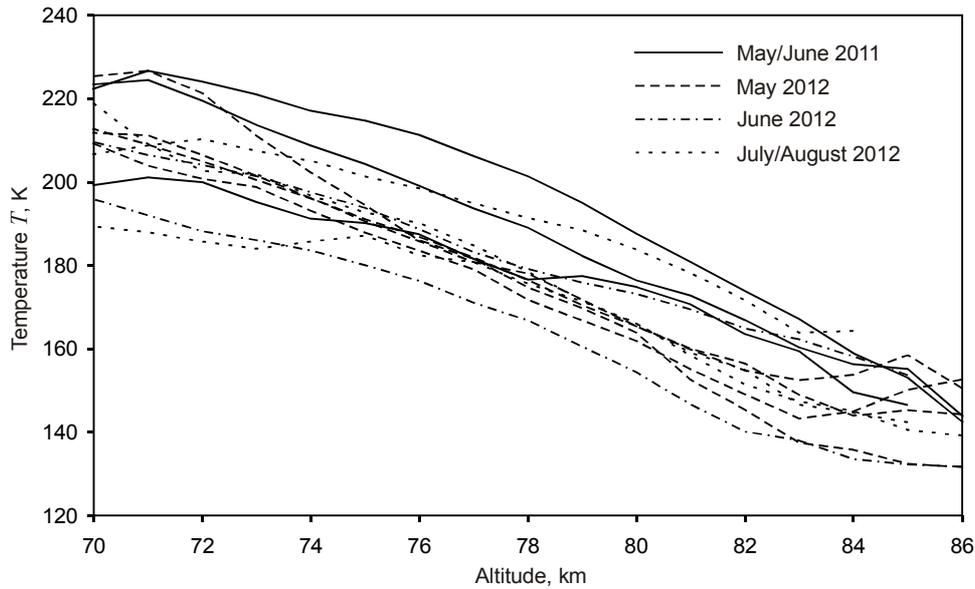

*Figure 12. Mesosphere temperature profiles for some observed twilights in 2011 and 2012.*

We see that SABER and MLS temperature data are shifted by about 5K one from another, the effect was noticed by Schwartz et al. (2008) especially for northern hemisphere. However, this exceeds the 1σ value only at 70 km. The WAPC temperature estimations seem to be in good agreement with both satellites data, being (in average) in between. Finally, Fig.12 shows the temperature profiles for a number of twilights with the best accuracy at the whole upper mesosphere altitude range.

**5. Discussion and conclusion.**

The paper presents the results of frequent wide-angle polarization measurements of the sky during the whole twilight period (from sunset/sunrise to deep night) conducted in the summer during the solar activity maximum epoch. The basic summary of the polarization analysis is its use for simple and effective method of single scattering separation simultaneously testing the lower atmosphere conditions stability. The single scattering field is found for the altitude range 70-85 km and turned out to be Rayleigh-dominated for all observed twilights. Their list includes the evening twilight of August, 12, 2012, when the upper atmosphere was expected to contain Perseid meteoric dust, moderated there during the several days before. However, we have just one twilight data obtained during the meteor activity epoch and some depolarization effects below the noise level are observed. We had also to take into account that the effective altitude of meteoric dust moderation can exceed 85-90 km, and the Leonids outburst activity in 2002 (Arlt et al., 2002), when the depolarization effects were seen (Ugolnikov and Maslov, 2007) was more than 30 times stronger than usual Perseids activity in 2012.

Domination of Rayleigh scattering gives the possibility to find the temperature of upper mesosphere as a function of altitude. We see that the accuracy of such measurements and agreement with space experiments data are quite good. It is important that the work range of the method includes the summer mesosphere, where the minimal atmospheric temperatures (down to 130-140 K) are observed. It should be added that this kind of mesosphere temperature measurements is sufficiently less expensive than any other space or ground-based one. It can be the part of stationary or mobile (mass about 6 kg) complex of mesosphere temperature and aerosol (if appears) investigations.

**Acknowledgements**

Authors would like to thank Oksana Shalygina (Astronomical Observatory of Kharkiv State University, Ukraine) for the help with polarization camera calibration. The work is financially supported by Russian Foundation for Basic Research, grant №12-05-00501.